\documentclass[12pt,preprint]{aastex}

\shorttitle{Efficient Method for Evaluation of Secular Effects} 
\shortauthors{L. \'{A}. Gergely et al.} 

\begin{document}

\title{An Efficient Method for the Evaluation of Secular Effects \\
in the Perturbed Keplerian Motion}
\author{L\'aszl\'o \'A. Gergely, Zolt\'an Keresztes, Bal\'azs Mik\'oczi}
\affil{Departments of Theoretical and Experimental Physics, 
University of Szeged, 
Szeged 6720, Hungary}

\begin{abstract}
Binary systems subject to generic perturbations evolve on quasiperiodic
orbits. We derive the most generic class of perturbations, which allow to
evaluate secular effects via generalized complex true and eccentric anomaly
parameters, by use of the residue theorem. Such perturbations include both
the generic Brumberg force and various linear contributions to the
conservative dynamics of compact binaries, up to the second post-Newtonian
order. As an example, we compute the self-spin contribution to the
luminosity due to gravitational radiation of a compact binary consisting of
galactic black holes, the most important type of source for LISA. This
translates to simply calculate the residue of the instantaneous energy loss
in the origin of the complex parameter plane, which illustrates the power of
the presented method.
\end{abstract}


\section{Introduction}

The perturbed two-body problem is one of the central areas of interest of
Celestial Mechanics. A wide class of perturbations of the dynamics of a
binary system are described by the generic Brumberg force [\citet{Brumberg}
]. Another type of perturbation is provided by general relativistic
modifications of gravitational dynamics. Such measurable effects arise in
planetary motion (like the general relativistic contribution to the
precession of the perihelion of the planet Mercury), however more important
are the modifications in the strong-field regime, typically appearing in
compact binaries composed of neutron stars and / or black holes. A
well-known example is the Hulse-Taylor double pulsar [\citet{HT}], the
orbital period of which is continuously decreasing as gravitational
radiation escapes from the system [\citet{Taylor79}] (for a recent review
see [\citet{Taylor04}]). Efforts to directly detect gravitational waves with
Earth-based interferometric detectors (LIGO[\citet{LIGO}], VIRGO 
[\citet{VIRGO}], GEO [\citet{GEO}] and TAMA [\citet{TAMA}]) are currently under
way. Gravitational waves from distant coalescing galactic black hole
binaries will be searched for by the forthcoming Laser Interferometer Space
Antenna (LISA) [\citet{LISA}]. With improving detector sensitivity, upper
limits for the signal emitted by different configurations of gravitational
wave sources were already set ([\citet{LIGONS}], [\citet{LIGOBH}], [\citet{LIGOpulsars}]). 
Capturing gravitational waves in the noisy background
requires a 3.5 post-Newtonian (PN) accuracy both in the orbital phase [\citet{Blanchet}] 
and in the energy loss [\citet{Poisson95}] occurring in a
compact binary.

The backreaction of the escaping gravitational radiation on the orbit
decreases the orbital period of the compact binary. Although higher order
corrections to this effect are too small to be directly detected in the
early stages of the inspiral, they are important in the latter stages, when
the post-Newtonian parameter increases up to $0.1$ (at $10$ gravitational
radii separation) and higher order contributions become quite significant.
Therefore the contributions to the energy loss $<dE/dt>$ and angular
momentum loss $<dL/dt>$ have been worked out up to 2PN order accuracy in the
wave generation (or equivalently, to 4.5 PN order following the Keplerian
picture), in the case of eccentric orbits, and up to even higher orders for
circular orbits. (Here $<>$ denotes the time-average over one period, e.g. a
secular contribution.) The corrections arise at each PN order from both
general relativistic modifications of the dynamics and/or as a manifestation
of various physical characteristics of the system, like the spins, mass
quadrupole momenta and magnetic dipole momenta of the constituents. The
final aim of gravitational wave astronomy is to incorporate all of these
physical characteristics into accurate templates to be used in data
processing. Recovering the physical characteristics of astronomical objects
from a simultaneous analysis of data collected in both the electromagnetic
spectrum and gravitational wave bandwidth would be a major breakthrough in
modern astronomy.

Driven by this motivation, the interest in the perturbed two-body problem
has been renewed. In this context, in [\citet{param}] a generic perturbed
two-body problem with radial equation 
\begin{eqnarray}
\dot{r}^{2} &=&\dot{r}_{N}^{2}+\sum_{i=0}^{p}\frac{\varphi _{i}}{\mu
^{2}r^{i}}\ ,  \label{radial} \\
\dot{r}_{N}^{2} &=&2\frac{E}{\mu }+2\frac{Gm}{r}-\frac{L^{2}}{\mu ^{2}r^{2}}
\end{eqnarray}
was considered. The point mass $\mu $ (representing the reduced mass of the
binary) is orbiting a fixed gravitational centre $m$ (the total mass of the
system) at distance $r$ (the separation of the components). The energy $E$
and the magnitude of the orbital angular momentum $L$ are constants of the 
\textit{perturbed} motion\footnote{
Separating $E$ and $L$ from the rest of the coefficients of $r^0$ and 
$r^{-2}$ in the radial equation is advantageous, as it will enable us 
to express all results in terms of constants with physical interpretation.}.
The expression $\dot{r}_{N}^{2}$ 
(the subscript $N$ denoting Newtonian) has the familiary
Newtonian functional form, however $E$ and $L$ characterize the perturbed
motion. 
The perturbing terms contain the small coefficients $\varphi _{i}$,
assumed \textit{constants}. As will be shown in detail in Section 3, both
the generic perturbing Brumberg force, the first post-Newtonian order (1PN)
general relativistic corrections [\citet{DD}] and the spin-orbit (SO)
interaction for compact binaries [\citet{BOC}] fit into this scheme. The
generic equation (\ref{radial}) defines a radial period $T$ and turning
points $r_{{}_{{}_{\min }^{\max }}}$.

With these assumptions the integrals of the type 
\begin{equation}
I\left( \omega ,n\right) =\int_{0}^{T}\frac{\omega }{r^{2+n}}dt
\label{integral}
\end{equation}
(with constant $\omega $ and $n$ arbitrary integer) could be easily
evaluated by use of the \textit{residue theorem}. Important steps in the
proof of [\citet{param}] were the introduction of suitable generalized true
and eccentric anomaly parametrizations of the radial motion, $r\left( \chi
\right) $ and $r\left( \xi \right) $ as 
\begin{eqnarray}
\frac{2}{r} &=&\frac{1+\cos \chi }{r_{\min }}+\frac{1-\cos \chi }{r_{\max }}
\ ,  \label{true} \\
{2}r &=&(1+\cos \xi )r{_{\min }}+(1-\cos \xi )r{_{\max }}\ ,
\label{eccentric}
\end{eqnarray}
together with their complex counterpart $z=\exp \left( i\chi \right) $ and $
w=\exp \left( i\xi \right) $, respectively. The advantage of this method
over other methods for computing the secular integrals (\ref{integral}) is
its striking simplicity. As proven in [\citet{param}], in the majority of
physically relevant cases the only pole is in the origin. Extremely rarely,
a second pole (also given) occurs. The method was successfully applied in
[\citet{GPV}] for the evaluation of the averaged energy and angular momentum
losses due to the SO interaction for compact binaries on elliptic orbit, and
the results were in agreement with those of [\citet{RiethSchafer}], obtained
by a different method.

The more accurate study of the inspiral phase of compact binaries however
reveals other important contributions to be taken into account. These are
the perturbations due to the (a) spin-spin (SS) interaction [\citet{BOC}];
(b) the interaction of the magnetic dipole moments for neutron stars with
huge magnetic fields (DD) [\citet{IT}] and (c) the contribution arising from
the motion of one of the binary components (treated as a monopole) in the
quadrupolar field of the other (QM) [\citet{Poisson}]. The corresponding
secular contributions to the energy and angular momentum losses could be
computed [\citet{spinspin}], [\citet{mdipole}], and [\citet{quadrup}] by the
method presented in [\citet{param}], in spite of the radial equation being 
\textit{different} from Eq. (\ref{radial}), which rendered these computations 
outside the domain of validity of Theorems 1 and 2 of [\citet{param}]. 
Therefore the question arose, whether the results presented in [\citet{param}] 
would hold more generically.

In all above mentioned cases the radial equation was a generalization of Eq.
(\ref{radial}) with the coefficients $\varphi _{i}$ being \textit{periodic
functions} of the true anomaly $\chi $. Thus in this
paper we raise the question, whether the results of [\citet{param}] would
hold for \textit{generic} periodic functions $\varphi _{i}\left( \chi
\right) $.\ The answer we find is that some restrictions apply, and we
derive the class of admissible periodic functions $\varphi _{i}$. Section 2
contains the announcement and proof of the \textit{generalized} Theorems 1
and 2, as compared to [\citet{param}].

Starting from the explicit expressions for the respective interactions, we
show in Section 3, that the SS, QM and DD perturbations fit into
the class of admissible perturbing functions $\varphi _{i}$.

In Section 4, as a simple application of the presented method, we give the
self-interaction spin contributions to the luminosity of a compact binary
consisting of black holes, in the case when $S_{2}\ll S_{1}$.

\section{Admissible perturbations and the integration method}

The turning points of the radial motion (\ref{radial}), with $\varphi
_{i}=\varphi _{i}\left( \chi \right) $ are the loci $\dot{r}=0$, given by: 
\begin{equation}
r_{{}_{{}_{\min }^{\max }}}\!=\!\frac{Gm\mu \!\pm \!A_{0}}{-2E}\!\pm \!\frac{
1}{2\mu A_{0}}\!\sum_{i=0}^{p}\!\varphi _{i}^{\pm }\!\left[ \!\frac{\mu
(Gm\mu \!\mp \!A_{0})}{L^{2}}\!\right] ^{i-2}\ .  \label{rminmax}
\end{equation}
Here $A_{0}$ represents the magnitude of the Laplace-Runge-Lenz vector
characterizing a Keplerian motion with $E$ and $\ L$ and we have introduced
the notations 
\begin{equation}
\varphi _{i}^{-}=\varphi _{i}(0),\qquad \varphi _{i}^{+}=\varphi _{i}(\pi )\
.
\end{equation}
Unless in the case presented in [\citet{param}], here $\varphi _{i}^{-}\neq
\varphi _{i}^{+}$. The expression of the turning points allow to write up
the parametrizations (\ref{true}) and (\ref{eccentric}) in detail, whenever
needed.

The integrals (\ref{integral}) could be formally rewritten in terms of the
integration variable $\chi $ as 
\begin{equation}
I\left( \omega ,n\right) =\int_{0}^{2\pi }\frac{\omega }{r^{n}}\left( \frac{
1 }{r^{2}}\frac{dt}{d\chi }\right) d\chi \,.  \label{int1}
\end{equation}

\subsection{The case $n\geq 0$}

For $n\geq 0$ the factor $r^{n}$ in the denominator of the integrand in Eq.
( \ref{int1}) has the binomial expansion 
\begin{equation}
\left[ \frac{2}{r\left( \chi \right) }\right] ^{n}=\sum_{k=0}^{n}\left(
_{k}^{n}\right) \!\left( \!\frac{1\!+\!\cos \chi }{r_{min}}\right)
^{k}\!\left( \frac{1\!-\!\cos \chi }{r_{max}}\right) ^{n-k}\ .
\label{binomial}
\end{equation}
By applying the chain rule $dt/d\chi =\dot{r}^{-1}dr/d\chi $, with $dr/d\chi 
$ derived from the true anomaly para\-metri\-za\-tion (\ref{true}), and
inserting $\dot{r}^{-1}$ as the series expansion of the radial equation (\ref
{radial}), we find for the second factor of the integrand (\ref{int1}) 
\begin{eqnarray}
\frac{1}{r^{2}}\frac{dt}{d\chi } &=&\frac{\mu }{L}\left( 1+\frac{L^{2}}{4\mu
^{2}A_{0}^{2}\sin ^{2}\chi }\Gamma \right) \,,  \label{dtdchi} \\
\Gamma &=&\sum_{i=0}^{p}\left( \Delta _{+}^{i}+\Delta _{-}^{i}\cos \chi - 
\frac{2\varphi _{i}\left( \chi \right) }{r^{i}}\right) \ ,  \label{Gamma} \\
\Delta _{\pm }^{i}\! &=&\!\frac{\mu ^{i}}{L^{2i}}\bigl[\varphi
_{i}^{-}(Gm\mu +A_{0})^{i}  \nonumber \\
\qquad &&\pm \varphi _{i}^{+}(Gm\mu -A_{0})^{i}\bigr]\ .
\end{eqnarray}
The second term of the expression (\ref{dtdchi}) contains $\sin ^{2}\chi $
in the denominator, and apparently becomes singular at $\chi =k\pi $ ($k$
any integer). The key point of the proof of Theorem 1 of [\citet{param}] was
that a factor of $\sin ^{2}\chi $ could be also separated from $\Gamma $, so
that the integrand of $I\left( \omega ,n\right) $ became regular everywhere.
The proof has also showed that there is no other $\chi $-dependence left in
the denominator.

We would like to see whether this convenient property of $\Gamma $ holds for
generic periodic functions $\varphi _{i}\left( \chi \right) $. For this we
decompose $\varphi _{i}\left( \chi \right) $ as follows 
\begin{equation}
\varphi _{i}\left( \chi \right) =\sum_{j=0}^{\infty }\left(
f_{ij}+g_{ij}\cos \chi \right) \sin ^{j}\chi \ ,  \label{series_phi}
\end{equation}
the coefficients $f_{ij}$ and $g_{ij}$ being constants. Then the condition 
\begin{equation}
\varphi _{i}^{\mp }=f_{i0}\pm g_{i0}  \label{fi0gi0}
\end{equation}
follows. Eq. (\ref{series_phi}) is equivalent with a decomposition into a
generic Fourier series, however better suited for our purposes. Its
advantage lies in the property of the terms with $j\geq 2$, of manifestly
containing the factor $\sin ^{2}\chi $. Therefore singular integrands in 
(\ref{int1}) could arise only from the terms containing $f_{i0}$, $g_{i0}$, $
f_{i1}$ and $g_{i1}$. We will discuss these one by one.

As\ $f_{i0}$ is the constant part of $\varphi _{i}\left( \chi \right) $, the
contributions containing $f_{i0}$ are the constant perturbing functions
already dealt with in [\citet{param}]. Thus the terms containing $f_{i0}$
are regular.

A similar algebra as for the terms with $f_{i0}$ applies to the terms
containing $g_{i0}$. By employing the binomial expansion (\ref{binomial})
and the expression of the turning points (\ref{rminmax}), those
contributions to $\Gamma $ which contain the factor $2\left( \mu
/L^{2}\right) ^{i}g_{i0}$ sum up to 
\begin{equation}
\sum_{l=0}^{i}\left( _{l}^{i}\right) \left( Gm\mu \right) ^{i-l}A_{0}^{l} 
\left[ \frac{1-\left( -1\right) ^{l}}{2}\sin ^{2}\chi +\Xi _{l+1}\left( \chi
\right) \right] \,,
\end{equation}
The expressions $\Xi _{k}\left( \chi \right) $ are defined by Eqs. (29)-(31)
of [\citet{param}], and they are proportional to $\sin ^{2}\chi $.\ This
proves that the terms with $g_{i0}$ do not give singular contributions to
the integrand of $I\left( \omega ,n\right) $.

What remains to check are the $j=1$ terms of $\Gamma $, given by 
\begin{equation}
-\sum_{i=0}^{p}\frac{2\left( f_{i1}+g_{i1}\cos \chi \right) \sin \chi }{
r^{i} }\,.  \label{Gamma1}
\end{equation}
(We have used that, according to Eq. (\ref{fi0gi0}), the expressions $\Delta
_{\pm }^{i}$ receive no contribution from $f_{i1}$ and $g_{i1}.$) \ As the
coefficients $f_{i1},\ g_{i1}$ are already of first order, the Keplerian
true anomaly parametrization $r=\left( \mu /L^{2}\right) \left( Gm\mu
+A_{0}\cos \chi \right) $ can be employed in Eq. (\ref{Gamma1}).
A tedious but straightforward algebra, based on (a) the binomial expansion
of $1/r^{i}$; (b) the replacement cos$^{2}\chi =1-\sin ^{2}\chi $, whenever
possible; (c) a second binomial expansion of $(1-\sin ^{2}\chi )^{q}$,
whenever applicable and (d) grouping together the terms without a $\sin
^{2}\chi $ factor, into terms proportional to $\sin \chi $ and $\sin \chi
\cos \chi $, leads to the following two conditions on the coefficients $
f_{i1}$ and $g_{i1}$: 
\begin{equation}
\sum_{i=0}^{p}\frac{\mu ^{i}}{L^{2i}}\left( Gm\mu \pm A_{0}\right)
^{i}\left( f_{i1}\pm g_{i1}\right) =0\ ,  \label{condition1}
\end{equation}
which have to be satisfied in order $dt/d\chi $ to be regular for every $
\chi $.

The last question to be addressed is what are the conditions to be imposed
on $\omega $, in order that $I\left( \omega ,n\right) $ stays regular. The
answer is simple: $\omega $ can be any periodic function of $\chi $, which
is regular. Such functions will not affect the proof presented above.

With these conditions satisfied, all factors in the integrand of $I\left(
\omega ,n\right) $ become simple polynomials in $\sin \chi =\left(
z^{2}-1\right) /2iz$ and $\cos \chi =\left( z^{2}+1\right) /2z$, provided $
n\geq 0$. This renders the only pole in the origin, and we have proven:

\textbf{Theorem 1}: \textsl{\ For all perturbed Keplerian motions
characterized by the radial equation (\ref{radial}), with periodic
perturbing functions }$\varphi _{i}$\textsl{\ obeying the conditions (\ref
{series_phi}) and (\ref{condition1}),\ and for arbitrary periodic functions }
$\omega $\textsl{\ the integral }$I\left( \omega ,n\geq 0\right) $ \textsl{\
is given by the residue in the origin of the complex true-anomaly parameter
plane. }

\subsection{The case $n<0$}

In order to cover the complementary case of $n<0$, we employ the eccentric
anomaly parametrization (\ref{eccentric}). The integrals $I\left( \omega
,n\right) $ then are evaluated as 
\begin{equation}
I\left( \omega ,n\right) =\int_{0}^{2\pi }\frac{\omega }{r^{n+1}}\left( 
\frac{1}{r}\frac{dt}{d\xi }\right) d\xi \,,
\end{equation}
and for any $n^{\prime }\equiv -n-1\geq 0$ we apply the binomial expansion 
\begin{equation}
\left( {2}r\right) ^{n^{\prime }}=\sum_{k=0}^{n^{\prime }}\left( _{n^{\prime
}}^{k}\right) r_{\min }^{k}r_{\max }^{n^{\prime }-k}(1+\cos \xi )^{k}(1-\cos
\xi )^{n^{\prime }-k}\ ,
\end{equation}
which is a polynomial in $\cos \xi $. From the eccentric anomaly
parametrization (\ref{eccentric}) and the radial equation (\ref{radial})
(with $\varphi \left( \chi \left( \xi \right) \right) $ in place of $\varphi
\left( \chi \right) $) we find 
\begin{eqnarray}
\frac{1}{r}\frac{dt}{d\xi } &=&\sqrt{\frac{\mu }{-2E}}\left( 1-\frac{E}{2\mu
A_{0}^{2}\sin ^{2}\xi }\Upsilon \right) \ ,  \label{dtdxi} \\
\Upsilon &=&\sum_{i=0}^{p}\left( \Omega _{+}^{i}-\Omega _{-}^{i}\cos \xi - 
\frac{2\varphi _{i}\left( \xi \right) }{r^{i-2}}\right) \ ,  \label{iota} \\
\Omega _{\pm }^{i}\! &=&\!\left( \frac{\mu }{L^{2}}\right) ^{i-2}\!\bigl[ 
\varphi _{i}^{+}(Gm\mu \!-\!A_{0})^{i-2}\!  \nonumber \\
&&\qquad \qquad \pm \!\varphi _{i}^{-}(Gm\mu \!+\!A_{0})^{i-2}\!\bigr]\ .
\end{eqnarray}
Here $\varphi _{i}^{-}$ and $\varphi _{i}^{+}$ are identical with those
defined in terms of the true anomaly (this is, because at $r_{{}_{{}_{\min
}^{\max }}}\,$both angles $\chi $ and $\xi $ take the values $0$ and $\pi $,
respectively)$.$ Considerations completely analogous to those in the proof
of Theorem 1, together with the relation between the two parametrizations 
\begin{eqnarray}
\cos \chi &=&\frac{Gm\mu \cos \xi -A_{0}}{Gm\mu -A_{0}\cos \xi }\ , \\
\sin \chi &=&\frac{(-2EL^{2}/\mu )^{1/2}\sin \xi }{Gm\mu -A_{0}\cos \xi }\ ,
\end{eqnarray}
after lengthy computations lead to the following result. Provided the Eqs. 
(\ref{condition1}) hold, a factor of $\sin ^{2}\xi $ separates out from $
\Upsilon $ and cancels out the $\sin ^{2}\xi $ factor of the denominator in
Eq. (\ref{dtdxi}).

As main difference with respect to the true anomaly parametrization, the
terms $\varphi _{i}\left( \xi \right) /r^{i-2}$ in the expression (\ref{iota}
) could lead to additional factors $\left( Gm\mu -A_{0}\cos \xi \right) $ in
the denominator of the integrand. Therefore, the second pole 
\begin{equation}
w_{1}=\left( \frac{Gm\mu ^{2}-\sqrt{-2\mu EL^{2}}}{Gm\mu ^{2}+\sqrt{-2\mu
EL^{2}}}\right) ^{1/2}  \label{w1}
\end{equation}
could appear in the process of passing to a complex eccentric anomaly
parameter. We can now enounce our

\textbf{Theorem 2}: \textsl{For all perturbed Keplerian motions
characterized by the radial equation (\ref{radial}), with periodic
perturbing functions }$\varphi _{i}$\textsl{\ obeying the conditions (\ref
{series_phi}) and (\ref{condition1}),\ and for arbitrary periodic functions }
$\omega $,\textsl{\ the integral }$I\left( \omega ,n<0\right) $ \textsl{is
given by the sum of the residues in the origin of the complex
eccentric-anomaly parameter plane and (for any }$f_{ij}$ \textsl{with\ }$
i+j>2$\textsl{\ or }$g_{ij}$\textsl{\ with }$i+j>1$\textsl{\ in }$\varphi
_{i}$\textsl{) at }$w_{1}$\textsl{.}

Our Theorems contain as special cases the Theorems presented in 
[\citet{param}]. There, the conditions (\ref{condition1}) are automatically
satisfied, as $\varphi _{i}$ are constants. It is remarkable, that the
constancy of $\varphi _{i}$ can be relaxed in a quite generic way, in which
the infinite number of coefficients in the expansion of $\varphi _{i}\left(
\chi \right) $ should obey only the two contraints (\ref{condition1}).

\section{Admissible dynamical systems}

In this Section we present two important classes of dynamical systems, for
which our Theorems apply.

\subsection{Perturbations characterized by the generic Brumberg force}

The generic perturbing Brumberg force [\citet{Brumberg}], [\citet{Soffel}],
includes a wide range of perturbations employed in Celestial Mechanics. It
can be derived from the Lagrangian 
\begin{eqnarray}
\mathcal{L}_{B}&=&\mathcal{L}_{N}+\mathcal{L}_{PB} \\
\mathcal{L}_{N}&=&\frac{\mu \mathbf{v}^{2}}{2}+\frac{Gm\mu }{r} \ , \\
\mathcal{L}_{PB}&=&\frac{1}{4c^{2}}\left( \alpha -\beta +\frac{\lambda }{2}
\right) \mu \mathbf{v^{4}}+\alpha \frac{Gm\mu }{c^{2}r^{3}}(\mathbf{rv})^{2}
\nonumber \\
&&+\left( \beta -\alpha +\frac{\lambda }{2}\right) \frac{Gm\mu }{c^{2}r}
\mathbf{v}^{2}+\left( \beta -\gamma +\frac{\lambda }{2}\right) \frac{
G^{2}m^{2}\mu }{c^{2}r^{2}}\ .  \label{LagB}
\end{eqnarray}
where $\alpha ,\beta ,\gamma $ and $\lambda $ are perturbation parameters.
The radial equation derived in [\citet{param}] from this Lagrangian is of
the form (\ref{radial}), with the small coefficients 
\begin{eqnarray}
c^{2}\varphi _{0}^{B} &=&-3(2\alpha -2\beta +\lambda )E^{2}\,,  \nonumber \\
c^{2}\varphi _{1}^{B} &=&-4Gm\mu (3\alpha -2\beta +2\lambda )E\,,  \nonumber \\
c^{2}\varphi _{2}^{B} &=&-2\mu ^{2}G^{2}m^{2}(3\alpha -2\beta +2\lambda
+\gamma )  \nonumber \\
&&+\frac{2}{\mu }(2\alpha -2\beta +\lambda )EL^{2}\,,  \nonumber \\
c^{2}\varphi _{3}^{B} &=&2Gm(\alpha +2\lambda )L^{2}\ .
\label{radialBrumberg}
\end{eqnarray}
As all $\varphi _{i}^{B}$ are constants, these perturbations fit into the
domain of validiy of the methods discussed in [\citet{param}].

\subsection{Orbital evolution of compact binaries}

Another important example is provided by compact binary systems, consisting
of black holes / neutron stars. Their dynamics is conservative up to the 2PN
order, the first dissipative effects due to gravitational radiation occuring
at 2.5 PN order. The conservative dynamics including all leading order
contributions, due to either various physical characteristics, or to the
general relativistic modifications of the motion is described by the
Lagrangian 
\begin{equation}
\mathcal{L}_{CB}=\mathcal{L}_{N}+\mathcal{L}_{PN}+\mathcal{L}_{SO}+\mathcal{L
}_{SS}+\mathcal{L}_{QM}+\mathcal{L}_{DD}\ ,
\end{equation}
with the various contributions derived first in [\citet{DD}] (PN), 
[\citet{Kepler}], (SO), [\citet{KWW}] (SS), [\citet{Poisson}] (QM) and 
[\citet{IT}] (DD) read\footnote{
The magnitude and direction of the spins are denoted as $S_{i}$ and $\mathbf{
\ \hat{S}}_{\mathbf{i}}$. The angle subtended by them is $\gamma =\cos ^{-1}(
\mathbf{\hat{S}_{1}\cdot \hat{S}_{2})}$. The total spin is $\mathbf{S}=
\mathbf{S}_{\mathbf{1}}+\mathbf{S}_{\mathbf{2}}$ and ${
\mbox{\boldmath 
$\sigma$}=}\left( m_{2}/m_{1}\right) \mathbf{S}_{\mathbf{1}}+\left(
m_{1}/m_{2}\right) \mathbf{S}_{\mathbf{2}}$. The magnitude and direction of
the magnetic dipole moments $\mathbf{d}_{\mathbf{i}}$ are denoted as $d_{i}$
and $\mathbf{\hat{d}_{i}}$. They subtend the angle $\lambda =\cos ^{-1}(
\mathbf{\hat{d}_{1}\cdot \hat{d}_{2})}$ with each other.\ In a coordinate
systems $\mathcal{K}$ with the axes $(\mathbf{\hat{c},\hat{L}\times \hat{c},
\hat{L}})$, where $\mathbf{\hat{c}}$ is the unit vector in the $\mathbf{\
J\times L}$ direction, the polar angles $\kappa _{i}$ and $\psi _{i}$ of the
spins are defined as $\mathbf{\hat{S}_{i}=}(\sin \kappa _{i}\cos \psi
_{i},\sin \kappa _{i}\sin \psi _{i},\cos \kappa _{i})$ (see [\citet{GPV}]).
In the coordinate system $\mathcal{K}^{i}$ with the axes $(\mathbf{\hat{b}
_{i},\hat{S}_{i}\times \hat{b}_{i},\hat{S}_{i}})$, where $\mathbf{\hat{b}_{i}
}$ are the unit vectors in the $\mathbf{S_{i}\times L}$ directions,
respectively, the polar angles $\alpha _{i}$ and $\beta _{i}$ of the the
magnetic dipole moments $\mathbf{{d}_{i}\,}$are $\mathbf{{\hat{d}}_{i}=}
(\sin \alpha _{i}\cos \beta _{i},\sin \alpha _{i}\sin \beta _{i},\cos \alpha
_{i})$ (see [\citet{mdipole}]). The quadrupolar parameters (see 
[\citet{quadrup}]) are defined as $p_{i}=Q_{i}/m_{i}m^{2}$, where $Q_{i}$ is
the quadrupole-moment scalar [\citet{Poisson}] of the $\ i^{th}$ axially
symmetric binary component with symmetry axis $\mathbf{\hat{S}}_{\mathbf{i}}$
.}: 
\begin{eqnarray}
\mathcal{L}_{PN} &=&\frac{1}{8c^{2}}\left( 1-3\eta \right) \mu v^{4}+\!\frac{
Gm\mu }{2rc^{2}}\!\left[ \!\left( 3\!+\!\eta \right) v^{2}\!+\!\eta \dot{r}
^{2}\!-\!\frac{Gm}{r}\!\right]  \ ,  \nonumber \\
\mathcal{L}_{SO} &=&\frac{G\mu }{2c^{2}r^{3}}\mathbf{v}\cdot \lbrack \mathbf{
\ r}\times (4\mathbf{S}+3{\mbox{\boldmath $\sigma$}})] \ ,  \nonumber \\
\mathcal{L}_{SS} &=&\frac{G}{c^{2}r^{3}}\left[ \left( \mathbf{S}_{\mathbf{1}
}\cdot \mathbf{S}_{\mathbf{2}}\right) -\frac{3}{r^{2}}\left( \mathbf{r\cdot S
}_{\mathbf{1}}\right) \left( \mathbf{r\cdot S}_{\mathbf{2}}\right) \right] 
 \ ,  \nonumber \\
\mathcal{L}_{QM} &=&\frac{G\mu m^{3}}{2r^{5}}\sum_{i=1}^{2}p_{i}\left[
3\left( \mathbf{\hat{S}}_{\mathbf{i}}\cdot \mathbf{r}\right) ^{2}-r^{2}
\right]  \ ,  \nonumber \\
\mathcal{L}_{DD} &=&\frac{1}{r^{3}}\left[ 3(\mathbf{n\cdot d_{1}})(\mathbf{\
n\cdot {d}_{2}})-\mathbf{d_{1}\cdot {d}_{2}}\right]  \ .  \label{Lag}
\end{eqnarray}

Note that the 1PN accurate relativistic corrections to the Keplerian motion
characterized by $\mathcal{L}_{PN}$ is the particular case of the perturbing
Brumberg Lagrangian $\mathcal{L}_{PB}$, with the specifications $\alpha
=\eta /2,\beta =(1+3\eta )/2,\gamma =2+\eta ,\lambda =2-\eta $ (where $\eta
=\mu /m$). \ Therefore the PN perturbations fit into the domain of validiy
of the methods discussed in [\citet{param}].

The same holds for the SO perturbations, where the only non-vanishing $
\varphi _{i}^{SO}$ in the radial equation is 
\begin{equation}
\varphi _{3}^{SO}=-\frac{G\mu }{c^{2}}\left( 4\mathbf{L\cdot S}+3{\ 
\mbox {\boldmath {\bf 
L}$\cdot \sigma$}}\right) \ .  \label{phiSO}
\end{equation}
This is a constant, as the angles $\kappa _{i}$ span by the spins with the
orbital angular momentum $\mathbf{L}$ are constants (to leading order) 
\footnote{
Note however that neither $\mathcal{L}_{SO}$ nor the radial equation with
the perturbing function (\ref{phiSO}) characterize uniquely the
SO-perturbation. This is, because a spin-supplementary condition (SSC)
should be additionally imposed. We have followed here the SSC of 
[\citet{SSC1}] and [\citet{SSC2}]. There are other known SSC-s, including the
covariant SSC employed in [\citet{Kidder}], which gives the non-vanishing
constant perturbations $\varphi _{2}^{SO,cov}={\ \left( 2E/c^{2}m\right) ({
\mbox {\boldmath {\bf L}$\cdot \sigma$}})}\ $and $\varphi
_{3}^{SO,cov}=-\left( 2G\mu /c^{2}\right) (2\mathbf{L\cdot S}+{\ 
\mbox {\boldmath {\bf 
L}$\cdot \sigma$}})$. These can again be expressed in terms of the angles $
\kappa _{i}\,$, which are cons$\tan $ts in the leading order, required
here.{}}.

The situation is entirely different for the SS, QM and DD contributions to
the dynamics of a compact binary. In all these cases the energy $E$ is a
constant of motion, while (in contrast with the 1PN and SO contributions)
the magnitude $\ L$ of the orbital angular momentum is not. Indeed, with the
notable exception of the PN contribution, it can be derived from the
respective Lagrangians that the orbital angular momentum $\mathbf{L\equiv
r\times p}$ is not conserved:$\ \mathbf{\dot{L}}\neq 0$. While for SO
perturbations this means merely a precessional change about the vector $4
\mathbf{S}+3{\mbox {\boldmath $\cdot \sigma$}}$ (so that $L$ is conserved),
the situation is more complicated in the SS, QM and DD cases, in which $L$
is a function of the orbital position $L=L\left( \chi \right) $. In all
these cases however an angular average $\overline{L}$ over the radial motion
can be computed, which is conserved to the accuracy of linear effects we
consider here. The reason for this is that up to 2PN, the dynamics is
conservative. A description in terms of the constants of motion $E$ and $
\overline{L}$ is then in order, as described in detail in ([\citet{spinspin}
], [\citet{quadrup}] and [\citet{mdipole}]. The radial equation in all these
cases can be expressed in the form (\ref{radial}), with $\overline{L}$ in
place of $L$ while the perturbing terms contain the following explicit $\chi 
$-dependences: 
\begin{eqnarray}
\varphi _{2}^{SS} &=&\frac{G\mu ^{2}}{2c^{2}\overline{L}^{3}}S_{1}S_{2}\sin
\!\kappa _{1}\sin \!\kappa _{2}\{2\overline{A}\cos \left( \chi \!+\delta
\right) ]  \nonumber \\
&&+(3Gm\mu \!+\!2\overline{A}\cos \!\chi )\cos (2\chi \!+\delta )\}\ , 
\nonumber \\
\varphi _{3}^{SS} &=&\frac{G\mu }{c^{2}}S_{1}S_{2}[3\cos \kappa _{1}\cos
\kappa _{2}-\cos \gamma   \nonumber \\
&&-3\sin \kappa _{1}\sin \kappa _{2}\cos (2\chi \!+\delta )]\,,  \nonumber \\
\varphi _{2}^{QM} &=&-\frac{Gm^{3}\mu ^{3}}{2\overline{L}^{2}}
\!\!\sum_{i=1}^{2}p_{i}\sin ^{2}\!\kappa _{i}\{2\overline{A}\cos \!\left(
\chi \!+\!\delta _{i}\right)   \nonumber \\
&&+\!\left( 3G\mu m\!+\!2\overline{A}\cos \!\chi \right) \cos (\!2\!\chi
\!+\!\delta _{i})\}\,,  \nonumber \\
\varphi _{3}^{QM} &=&-G\mu ^{2}m^{3}\sum_{i=1}^{2}p_{i}\left[ 1-\!3\!\sin
^{2}\!\kappa _{i}\cos ^{2}\!\left( \!\chi \!+\frac{\delta _{i}}{2}\!\right) 
\right] \,,  \nonumber \\
\varphi _{2}^{DD} &=&-\frac{\mu ^{2}d_{1}d_{2}}{\overline{L}^{2}}\left[
(3Gm\mu +4\overline{A}\cos \chi )\mathcal{B}_{2}(\chi )-\overline{A}\sin
\chi \mathcal{B}_{2}^{\prime }(\chi )\right] \,,  \nonumber \\
\varphi _{3}^{DD} &=&-\mu d_{1}d_{2}\left[ \mathcal{A}_{0}-3\mathcal{B}
_{2}(\chi )\right] \,,  \label{phiVAR}
\end{eqnarray}
where $\!\psi _{0}$ is the angle between the periastron and the node line 
[\citet{GPV}], $\overline{\psi }=\left( \psi _{1}+\psi _{2}\right) /2$
represents an average, $\delta =2(\!\psi _{0}\!-\!\overline{\psi })$ and $
\delta _{i}=\!\psi _{0}\!-\!\psi _{i}$ differences in the azimuthal angles.
The quantty $\overline{A}$ is the magnitude of the Laplace-Runge-Lenz vector
characterizing a Keplerian motion with $E$ and $\ \overline{L}$. Finally, $
\mathcal{A}_{0}$ and $\mathcal{B}_{2}(\chi )$ denote 
\begin{eqnarray}
\mathcal{A}_{0} &=&2\cos \lambda +3(\rho _{1}\sigma _{2}-\rho _{2}\sigma
_{1})\sin \left( \delta _{1}-\delta _{2}\right)   \nonumber \\
&&-3(\rho _{1}\rho _{2}+\sigma _{1}\sigma _{2})\cos \left( \delta
_{1}-\delta _{2}\right)  \ , \\
\mathcal{B}_{2}(\chi ) &=&(\sigma _{1}\sigma _{2}-\rho _{1}\rho _{2})\cos
(2\chi +\delta )-(\rho _{1}\sigma _{2}+\rho _{2}\sigma _{1})\sin (2\chi
+\delta )\ ,
\end{eqnarray}
where $\rho _{i}=\sin \alpha _{i}\cos \beta _{i}$, $\sigma _{i}=\sin \alpha
_{i}\sin \beta _{i}\cos \kappa _{i}+\cos \alpha _{i}\sin \kappa _{i}$. Note
that for the SS, QM, and DD interactions there are no contributions $\varphi
_{i}$ for any $i\neq 2,\,3$.

It is straightforward to transform the expressions (\ref{phiVAR}) into the
form (\ref{series_phi}), and to identify the coefficients $f_{i1}$ and $
g_{i1}$. From among the various emerging contributions, those entering in
the constraint (\ref{condition1}) are $f_{21}=\overline{A}\gamma $, $
\,f_{31}=0\ $, $g_{21}=Gm\mu \gamma $ and $g_{31}=-\left( \overline{L}
^{2}/\mu \right) \gamma $, with $\gamma $ a complicated expression depending
on the particulars of the SS, DD or QM contributions 
\begin{eqnarray}
\gamma &=&\gamma _{SS}+\gamma _{QM}+\gamma _{DD} \ , \\
\gamma _{QM} &=&-\frac{6G\mu ^{3}m^{3}}{2\overline{L}^{2}}\sum_{j}p_{j}\sin
^{2}\kappa _{j}\sin 2\delta _{j} \ , \\
\gamma _{SS} &=&-\frac{6G\mu ^{2}S_{1}S_{2}}{c^{2}\overline{L}^{2}}\sin
\kappa _{1}\sin \kappa _{2}\sin \delta  \ , \\
\gamma _{DD} &=&\frac{6G\mu ^{2}d_{1}d_{2}}{\overline{L}^{2}}\biggl [(\sigma
_{1}\sigma _{2}-\rho _{1}\rho _{2})\sin \left( \delta _{1}+\delta _{2}\right)
\nonumber \\
&&+(\rho _{1}\sigma _{2}+\rho _{2}\sigma _{1})\cos \left( \delta _{1}+\delta
_{2}\right) \biggr ]\ \ .
\end{eqnarray}
It is easy to verify that the condition (\ref{condition1}) holds in all
cases, irrespective of the particular form of $\gamma $. This is why the
method developed originally for constant $\varphi _{i}$-s could be employed
for these cases as well

\section{The self-spin contribution to the luminosity of compact binaries}

Gravitational radiation produces dissipative effects in the orbital
evolution. While the energy $E$, and the total orbital momentum $\mathbf{J}$
are conserved up to 2PN order accuracy (including the conservative
perturbations considered in this paper and the 2PN relativistic perturbation), 
they start to evolve due to escaping gravitational radiation.
This dissipative evolution can be conveniently computed with our
method, presented in Section 2. The secular losses due to gravitational
radiation are exactly integrals of the type (\ref{integral}), in which the
condition of periodicity of $\omega $ is obeyed for all losses of 
SS, DD and QM type. As already argued in Section 3.b, the
radial equation written in terms of $\overline{L}$ is also of the form (\ref
{radial}) in all of these cases. Therefore our Theorems can be applied in
computing secular radiative changes. Such secular effects were discussed in
([\citet{spinspin}], [\citet{mdipole}] and [\citet{quadrup}]) for the SS, DD
and QM contributions, respectively.

In this Section we would like to apply the method for computing the
so-called self-spin contribution to the gravitational luminosity (energy
loss) of compact binaries composed of black holes. For a black hole the
quadrupole moment scalar is directly related to the spin: $
Q_{i}=-S_{i}^{2}/m_{i}\ $[\citet{Poisson}]. Therefore the QM losses are
effectively self-spin contributions and add to those discussed in 
[\citet{spinspin}], [\citet{self}]. Such contributions become important
whenever the mass ratio of the compact binary is at least $\eta =0.1$, which
is frequently the case for colliding galactic black holes, which are the
most important source types to be searched for by LISA. In such cases, as $
S_{2}/S_{1}\propto \eta ^{2}$, we can safely ignore $S_{2}$. Therefore the
only SS interaction is the self-spin contribution. 
\begin{table}[tbp]
\caption{The coefficients $z_{ij}$ in the self-spin contribution to the
luminosity.}
\label{table}
\begin{center}
\begin{tabular}{crrrr}
\tableline\tableline$_{i}\backslash ^{j}$ & $0$ & $1$ & $2$ & $3$ \\ 
\tableline$0$ & $-539\,784$ & $-1229\,200$ & $-\allowbreak 657\,120$ & $
-47\,040$ \\ 
$1$ & $797\,076$ & $1807\,400$ & $959\,280$ & $67\,680$ \\ 
$2$ & $266\,615$ & $818\,986$ & $601\,812\allowbreak $ & $60\,600$ \\ 
\tableline &  &  &  & 
\end{tabular}
\end{center}
\end{table}

\begin{figure}[th]
\centering\includegraphics[width=0.55\linewidth]{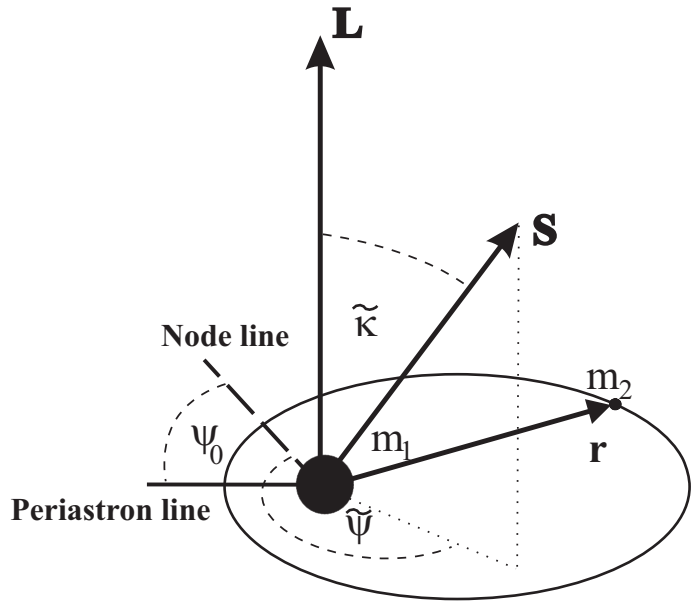}
\caption{The figure represents a black hole - black hole binary system. $
\mathbf{S}$ is the spin of the supermassive black hole with mass $m_{1}$,
while the spin of $m_{2}\ll m_{1}$ is negligible. As there is either no 
$SS$ or $QM$-contribution to the orbital angular momentum 
[\citet{Kidder}, \citet{quadrup}],  
$\mathbf{L}= \mu \mathbf{r}\times \mathbf{v}$ is the Newtonian orbital angular
momentum. $\widetilde{\kappa }$ is the angle between $\mathbf{S}$
and $\mathbf{L}$. The angle between the node line and
projection of the spin in the plane of the orbit is denoted as $\widetilde{ 
\psi }$. }
\label{Fig1}
\end{figure}

The instantaneous energy loss can be found by taking the appropriate
derivatives of the mass and current quadrupole tensors of the mass $\mu $,
as described for example in [\citet{Kidder}]. The self-spin contribution to
the instantaneous energy loss computed in this way is a complicated
expression of $r,\,\dot{r}$ and $\chi $. By employing the parametrization 
(\ref{true}), we obtain $dE/dt$ as function of $\chi $ alone. The secular
contribution to the energy loss arises by integration of $dE/dt$ 
over one radial orbit. This integration is carried out by computing the
residues enclosed in the circle $\zeta =e^{i\chi }$. Due to Theorem 1, we
are assured that there is only one pole, at $\zeta =0$. The summed-up
self-spin contribution to the luminosity is 
\begin{equation}
\mathcal{L}_{QM+SS}=-\frac{G^{2}(-2E\mu )^{3/2}S^{2}}{960c^{7}\overline{L}^{11}}
\frac{m_{2}}{m_{1}}\left\{ Z_{0}+\left[ Z_{1}+Z_{2}\cos 2\left( \psi _{0}-
\widetilde{\psi }\right) \right] \sin ^{2}\widetilde{\kappa }\right\} \ ,
\label{luminosity}
\end{equation}
where 
\begin{equation}
Z_{i}=G^{6}m^{6}\mu ^{9}\sum_{j=0}^{3}z_{ij}\left( \frac{E\overline{L}^{2}}{
G^{2}m^{2}\mu ^{3}}\right) ^{j}\ ,\quad i=0..2\,\ .  \label{Z}
\end{equation}
The angles $\widetilde{\kappa }$ and $\widetilde{\psi }$ are shown in Fig\ref
{Fig1}. The coefficients $z_{ij}$ are enlisted in table~\ref{table}. Eqs. 
(\ref{luminosity})-(\ref{Z}) contain the totality of second order
contributions in the spins to the luminosity, provided $S_{2}=0$, and as
such, represent the first correction to the Lense-Thirring approach,
discussed in [\citet{GPV}].

\section{Concluding Remarks}

Perturbed Keplerian motions can be para\-metr\-ized in many different ways.
(For a review of these parametrizations see [\citet{KlionerKopeikin}].) The
method of computing secular effects presented in [\citet{param}] relies on
novel complex parametrizations of the radial part of the motion and its
advantage over previous approaches [\citet{Ryan}], [\citet{RiethSchafer}] is
in its overwhelming simplicity. Other methods can of course be successfully
applied, like the use of the Laplace second integrals for the Legendre
polynomials ([\citet{WhittakerWatson}], [\citet{GopakumarIyer}]).

In the present work we have generalized both the class of integrands and the
class of perturbing forces for which the simple method relying on the use of
complex true and eccentric anomaly parametrizations applies.

In a generic perturbed two-body problem, it is easy to see whether a
particular secular effect to be computed is of the type (\ref{integral}),
with $\omega $ periodic, and then to check whether the periodic perturbing
functions $\varphi _{i}$\ obey the conditions (\ref{condition1}). Then,
depending on the power $n$, either Theorems 1 or 2 can be applied.

We have also shown explicitly, that for several physically
interesting linear perturbations in the conservative dynamics these Theorems
apply. Whether the SO, SS, QM or DD perturbations are of 1PN order or
higher, depends on the specific system considered. For comparable mass
compact binaries for example the SO contribution is of 1.5PN order, while
the SS, QM, DD contributions are all of order 2PN.

By contrast, when one of the masses of the compact binary dominates over the
other, the SS contribution becomes negligible, however the so-called
self-spin contribution of the dominant spin becomes important. Driven by
this remark, as an application to the presented method, we have computed the
self-spin contribution to the gravitational luninosity.

\section{Acknowledgments}

This work was supported by OTKA grants no. T046939 and TS044665. L.\'{A}.G.
wishes to thank the J\'{a}nos Bolyai Scholarship for support.

\end{document}